\documentclass[%
 aip,
 amsmath,amssymb,
 reprint,%
]{revtex4-1}

\usepackage{graphicx}
\usepackage{dcolumn}
\usepackage{bm}

\usepackage[utf8]{inputenc}
\usepackage[T1]{fontenc}
\usepackage{mathptmx}
\usepackage{etoolbox}

\makeatletter
\def\@email#1#2{%
 \endgroup
 \patchcmd{\titleblock@produce}
  {\frontmatter@RRAPformat}
  {\frontmatter@RRAPformat{\produce@RRAP{*#1\href{mailto:#2}{#2}}}\frontmatter@RRAPformat}
  {}{}
}%
\makeatother
\begin{document}

\preprint{AIP/123-QED}

\title[Sample title]{Voltage Noise Thermometry in Integrated Circuits at Millikelvin Temperatures }
\author{G. Ridgard*}
\email{g.ridgard@lancaster.ac.uk} 
\affiliation{ 
Lancaster University Physics Department, Lancaster, UK
}%
\affiliation{Quantum Motion Technologies, London, UK}
\author{M. Thompson}
\affiliation{ 
Lancaster University Physics Department, Lancaster, UK
}%
\author{L. Schreckenberg}
\affiliation{ 
Forschungszentrum Jülich GmbH, 52425 Jülich, Germany
}%
\author{N. Deshpande}
\affiliation{ 
Forschungszentrum Jülich GmbH, 52425 Jülich, Germany
}%
\author{A. Cabrera-Galicia}
\affiliation{ 
 Forschungszentrum Jülich GmbH, 52425 Jülich, Germany
}%
\author{O. Bourgeois}
\affiliation{ 
Institut NEEL, Univ. Grenoble Alpes, Grenoble, France
}%
\author{V. Doebele}
\affiliation{ 
Institut NEEL, Univ. Grenoble Alpes, Grenoble, France
}%
\author{J. Prance}
\affiliation{ 
Lancaster University Physics Department, Lancaster, UK
}%

\date{\today}

\begin{abstract}
This paper demonstrates the use of voltage noise thermometry, with a cross-correlation technique, as a dissipation-free method of thermometry inside a CMOS integrated circuit (IC). We show that this technique exhibits broad agreement with the refrigerator temperature range from 300~mK to 8~K. Furthermore, it shows substantial agreement with both an independent in-IC thermometry technique and a simple thermal model as a function of power dissipation inside the IC. As the device under test (DUT) is a resistor, it is feasible to extend this technique by placing many resistors in an IC to monitor the local temperatures, without increasing IC design complexity. This could lead to better understanding of the thermal profile of ICs at cryogenic temperatures. This has its greatest potential application in quantum computing, where the temperature at the cold classical-quantum boundary must be carefully controlled to maintain qubit performance.
\end{abstract}

\maketitle

\section{Introduction}
At deep cryogenic temperatures (< 4~K), power dissipation in electronic components can induce significant localised heating \cite{blagg2022chip}. The primary cause is the reduction in the thermal conductivity of both silicon \cite{glassbrenner1964thermal} and the insulating materials in its packaging \cite{chawner2019lego}, coupled with the increase of the thermal boundary resistances \cite{swartz1989thermal}. This issue is particularly important in Quantum Information Processing (QIP). The optimal performance of many types of qubit \cite{brandl2016cryogenic,corcoles2011protecting,veldhorst2017silicon} occurs at deep cryogenic temperatures due to diminished thermal noise. These qubits require ancillary classical electronics for control and read-out. As the qubit count scales, so does the requisite number of wires. To mitigate the unsustainable proliferation of wires into the cryostat, additional classical electronics are now co-located with the quantum components \cite{gonzalez2021scaling, pauka2021cryogenic}. Consequently, monitoring the local temperature at the qubit-classical electronics interface becomes crucial for maintaining consistent qubit operation. \\ While thermometry at deep cryogenic temperatures has been studied for many years, some existing solutions are inadequate for determining the temperature within an integrated circuit (IC). The most prevalent thermometry technique, temperature coefficient of resistance (TCR) thermometry \cite{courts2014standardized,courts2008commercial,dechaumphai2014sub}, is unsuitable because the thermometer will be poorly thermalised to the IC interior due to the low thermal conductivity of silicon and the thermal boundary resistance between the IC and the thermometer. Thermometry methods have been developed for use within ICs to address this issue \cite{huizinga2022integrated,noah2024cmos}. Methods such as Gate Resistance Thermometry (GRT) \cite{triantopoulos2019self,artanov2022self}, quantum dot thermometry \cite{chawner2021nongalvanic,de2024measurement,singh2024quantum}, and diode thermometry \cite{choi2021characterization,bebitov2023dependence} each have their individual merits and specific applications, but all have associated power dissipation inside the IC. Consequently, all will influence the internal temperature of the IC.\\ 
One such technique which is both dissipationless and primary is noise thermometry \cite{qu2019johnson}. This has been well established in a broad range of environments \cite{casey2014current,ivashchenko1975electron,brixy1996noise}.
However, it has several barriers to entry as a technique for measuring the internal temperature of an IC at 
temperatures that QIP platforms are located. The first problem is that in the case of voltage noise thermometry the device under test (DUT) resistor must have a very high resistance to make the noise measurable above the typical readout amplifier noise. 
High resistance values result in low cut off frequencies, making measurements extremely slow and susceptible to 1/f noise. This paper demonstrates the use of the cross-correlation technique to remedy this issue. Cross-correlation allows the use of lower value resistors by allowing the measurement of noise values below the noise of the amplifiers used in the setup. In this paper we demonstrate this by using cross-correlation to accurately measure the temperature of our DUT, a $\mathrm{45~kOhm}$ resistor, from 300~mK to 8~K. \\ Another problem in performing noise thermometry in an IC is the potential for non-thermal noise contributions. The resistors are typically made of doped silicon, which as a semiconductor has many more potential sources of noise \cite{wilamowski2018fundamentals} and the close proximity of power dissipating devices (on the order of $\mathrm{100~\mu m}$) may provide a source of cross-talk in the spectral measurements. In this paper, we resolve this problem by showing that even if the resistors have non-thermal noise sources, if this is appropriately characterised then its contribution can be subtracted from the spectral measurements, enabling accurate noise thermometry. We demonstrate this by first showing that our DUT has an elevated noise level at the fridge base temperature, which cannot be attributed to thermal noise. Next, from the base temperature measurement, we extract an offset noise (the non-thermal contribution). We then calibrate the noise temperature against the refrigerator temperature. It was found that when the offset noise is subtracted from the spectral measurements, the noise thermometry shows good agreement with the temperature of the refrigerator from 300~mK to 8~K. This was further tested by comparing the noise temperature against a different in-IC thermometry technique (TCR of the resistor) as a function of power. \\ In Sec II, we present our measurement setup and protocol of voltage noise thermometry inside an IC. Next, in Sec III, we examine the noise characteristics of the resistor selected for the experiment and its calibration against the temperature of the Mixing Chamber (MXC) in a refrigerator. Finally, we compare this technique to another internal thermometry method and a thermal model in Sec IV.
\section{Set up}
\begin{figure}
    \centering
    \includegraphics[width = 9cm]{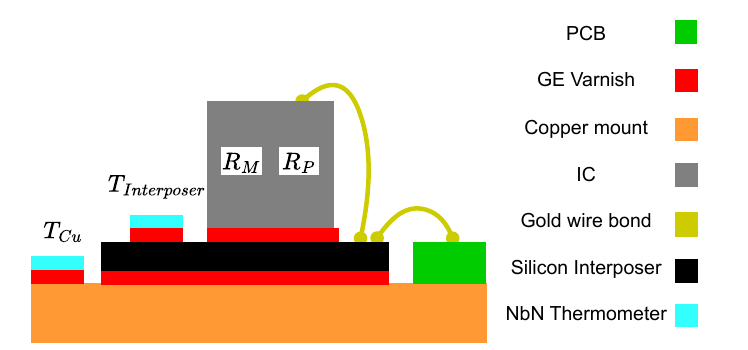}
    \caption{Illustration of the setup: PCB board composed of standard FR4. The GE varnish was lightly diluted to aid its application. The copper mount was constructed from a 6~mm thick piece of C110 oxygen free copper which has a minimum of $99.99~\%$ purity. The silicon interposer was $550 \mathrm{\mu m}$ thick and was lightly boron doped with gold patterned on top to allow for bonding between IC, interposer and PCB. Two Niobium Nitride (NbN) based thermometers were placed on the copper and silicon interposers to monitor their temperature as a function of power. Details on their operation and calibration can be found in Appendix C. This illustration is not to scale.  }
    \label{fig:Set_up_phys}
\end{figure}

Figure \ref{fig:Set_up_phys} illustrates the IC configuration. The IC, a 2~mm x 2~mm x 0.8~mm chip, is affixed to a silicon interposer, mounted on a ultra-high-purity copper mount, which in turn is secured to the puck connected to the MXC. The device is interconnected via gold bond wires to gold traces on the interposer, and subsequently bonded to the printed circuit board (PCB). The IC contains a $\mathrm{350~\Omega}$ power resistor, $R_{P}$, and a $\mathrm{45~k\Omega}$ measurement resistor, $R_{M}$. These components are fabricated using an industrial 65~nm bulk CMOS process. The entire assembly is cooled by an \textit{Oxford Instruments Triton 400} dilution refridgerator.
\subsection{Measurement Principle}
For a resistor with resistance $R_{M}$ at a temperature $T$, the Johnson-Nyquist theorem \cite{johnson1928thermal,nyquist1928thermal} states that the power spectral density, $S^{2}_{D}$, within a 1Hz bandwidth across the resistor is expressed as
\begin{equation}
    \label{eq:JN_voltage}
        S^{2}_{D} = 4k_{B}TR_{M} \quad \left[\frac{V^{2}}{Hz}\right].
\end{equation}
Consequently, if the resistance is precisely known, the resistor can serve as a primary thermometer, with the temperature determined by
\begin{equation}
    \label{eq:Temperature_cal}
    T = \frac{S^{2}_{D}}{4k_{B}R_{M}}.
\end{equation}To determine $S^{2}_{D}$, we calculate a cross-correlated spectrum to measure noise levels beneath the intrinsic noise of the amplifiers \cite{chen1965hanbury,wolff1965brown}. This demands the simultaneous acquisition of voltage as a function of time via two amplifier channels. The time traces of amplifiers $\alpha$ and $\beta$, denoted as $V_{\alpha}(t)$ and $V_{\beta}(t)$ respectively, contain a shared signal, $V_{C}(t)$, alongside a noise component pertinent to the input noise of each amplifier, $V_{\alpha:n}(t)$ and $V_{\beta:n}(t)$. In this context, $V_{C}(t)$ represents the thermal voltage originating from the resistor. The expressions for the amplifier signals are given by,
\begin{equation}
    V_{\alpha}(t) = V_{C}(t) + V_{\alpha:n}(t),
\end{equation}
and
\begin{equation}
    V_{\beta}(t) = V_{C}(t) + V_{\beta:n}(t).
\end{equation}Each cross-correlation spectrum, $S^{2}_{cc}$, is derived by computing the dot product of the Fourier transform of one trace with the complex conjugate of the Fourier transform of the other,
\begin{equation}
S^{2}_{cc}(f) \rightarrow \tilde{V}_{\alpha}(t)^{*} \cdot \tilde{V}_{\beta}(t).
\end{equation}
The resultant spectrum are then averaged $n$ times to attenuate uncorrelated noise, yielding $\langle S^{2}_{cc}(f) \rangle_{n}$. With adequate averaging, this converges to the Fourier transform of the common signal,
\begin{equation}
\label{eq:CC} 
\langle S^{2}_{cc}(f) \rangle_{n} = \tilde{V}_{C}(f).
\end{equation} 
Given the complex nature of this spectra, its magnitude is computed and subsequently scaled to account for the windowing effect during the Fourier transform. Finally, the average value over a frequency range $f_{1}-f_{2}$ encompassing $m$ points is calculated to produce our final value,
\begin{equation}
S^{2}_{D} = \frac{1}{m}\sum^{f_{2}}_{i=f_{1}}|\langle S^{2}_{cc}(f_{i}) \rangle_{n}|
\end{equation}
which is then converted to an Equivalent Noise Temperature (ENT) by Eq.\ref{eq:Temperature_cal}.
\subsection{Detail of operation}
\begin{figure*}
    \label{fig:Exp_setup}
    \centering
    \includegraphics[width = 18cm]{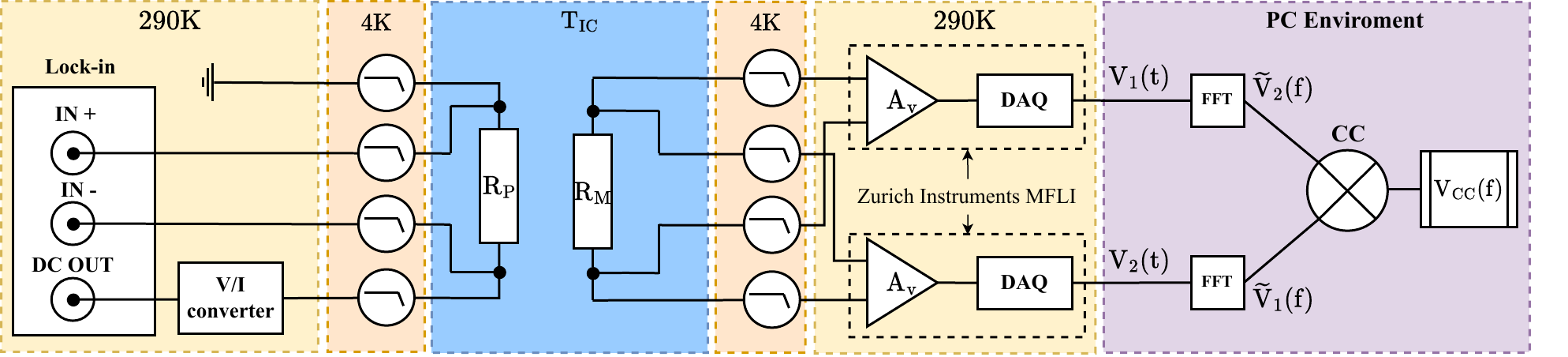}
    \caption{Experiment Schematic: The first box on the left side contains a \textit{Zurich Instruments MFLI} lock-in amplifier which provides a DC voltage through the auxiliary output to a \textit{Lancaster University}-designed low noise voltage-current converter, which converts at $\mathrm{1\mu A/V}$. This drives a current through the power resistor $R_{P}$. The same MFLI then measures the corresponding voltage response by demodulating at 0~Hz. The power through $R_{P}$ and as such in the IC is then given by the product of the DC current and DC voltage response. All lines down to low temperature are filtered at the 4~K plate by \textit{Aivon} Low Pass Filters (LPF). The next box shows the power resistor $R_{P}$ and the measurement resistor $R_{M}$ in the IC at a temperature $\mathrm{T_{IC}}$. The resistors are separated by $\mathrm{221~\mu m}$ in the IC. Finally, $R_{M}$ is connected to two MFLIs that provide simultaneous voltage readings across the resistor. Finally, both voltage time series are then Fourier transformed and have the cross-spectrum calculated by a Python script. }
    \label{fig:electrical_set_up}
\end{figure*}
A diagram of the electrical set up of the experiment is given in Fig. 2. The time traces in our experiment were recorded using two \textit{Zurich Instruments MFLIs} . Both amplifiers were synchronized and triggered via the Multi-Device Synchronization (MDS) package. To ensure the simultaneity of the traces, measurement timestamps from each amplifier were compared, ensuring that the traces commenced within a maximum of 20~$\mu s$. The amplifiers acquired data at $f_{clock}/2^{9} \approx$ 120~kHz, where $f_{clock}$ denotes the clock frequency of 60~MHz, collecting $\mathrm{2^{16}}$ samples. This configuration results in a measurement duration of 0.56~s per spectrum. Following the acquisition of both time traces, the cross-correlation spectrum was computed to attenuate non-correlated noise. This spectrum was subsequently averaged to further reduce residual noise. In our study, the number of averages was set to 3000, a value selected to limit the total measurement time to under 30 minutes. The requisite number of averages to sufficiently attenuate the measurement-specific noise floor is a function of both the noise, $\alpha$, (in $\mathrm{V/\sqrt{Hz}}$) of a single channel amplifier, and the noise floor targeted for resolution, $\gamma$. The number of averages to achieve a ratio of unity between the expected standard deviation of the measured value and the expected measured value is determined by
\begin{equation}
\label{eq:num_averages}
    n = \frac{1}{2\gamma^{4}}\left[\left(\alpha^{2} + \gamma^{2}\right)^{2} + \gamma^{4}\right]
\end{equation} 
which follows Rubiola \cite{rubiola20061} but conserves the contribution of the amplifier noise. Considering the number of averages and that at low frequencies the noise floor of a single channel is approximately $\mathrm{4~nV/\sqrt{Hz}}$, we can re-arrange Eq.\ref{eq:num_averages} to derive the minimum measurable noise floor,
\begin{equation}
    \label{eq:min_noise_floor}
    \gamma^{2} = \frac{\alpha^{2}}{\sqrt{2n-1} -1}.
\end{equation}
This yields $\gamma \approx \mathrm{0.5~nV/\sqrt{Hz}}$, corresponding to a minimum measurable temperature of approximately 100~mK (for $R = R_{M} = \mathrm{45~k\Omega}$). Conversely, in the absence of cross-correlation, the minimum resolvable noise would be that of the amplifier at low frequencies, with an equivalent minimum temperature of $\approx$ 6.4~K. Five distinct frequency ranges were selected to compute an average spectral density. These ranges were determined by examining the base temperature spectra and identifying regions without frequency peaks generated from interference from other electrical sources.
\subsection{Characterisation of measurement resistor}
\begin{figure}
    \centering
    \includegraphics{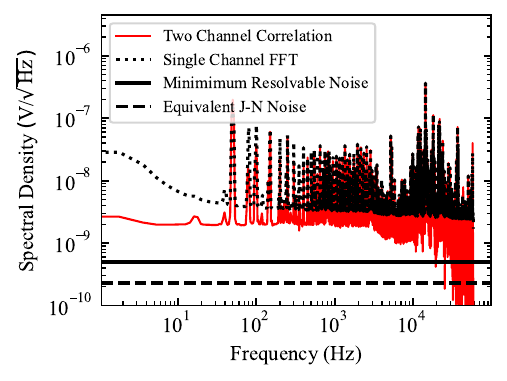}
    \caption{Base Temperature Measurement of $R_{M}$ after 3000 spectral averages. We see that the correlation is working as the two channel correlation (red line) is below that of the single channel FFT (dotted line). However, the two channel correlation should be much closer to the minimum resolvable noise (black line) as the thermal contribution at base (dashed line) is below the minimum resolvable noise. Therefore there must be another source of noise, other than the thermal contribution.}
    \label{fig:Initial_noise_measurement}
\end{figure}

\begin{figure}[h!]
    \centering
    \includegraphics{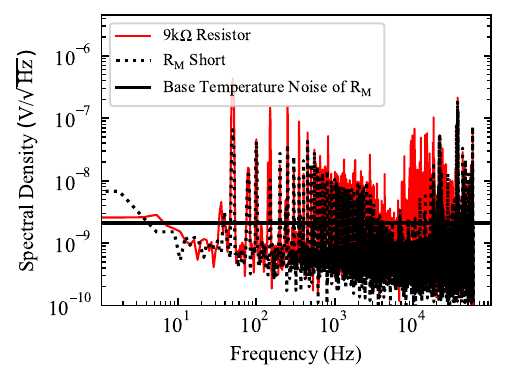}
    \caption{Base Temperature measurements of $R_{M}$ shorted and a $\mathrm{9~k\Omega}$ resistor: Here we see the two channel correlation of both the shorted $R_{M}$ and the $\mathrm{9~k\Omega}$ resistor are much below the noise floor of the base temperature measurement of $R_{M}$. This rules out the offset noise associated with $R_{M}$ being related to the set up. }
    \label{fig:Spectral_differences}
\end{figure}

\begin{figure}
    \centering
    \includegraphics{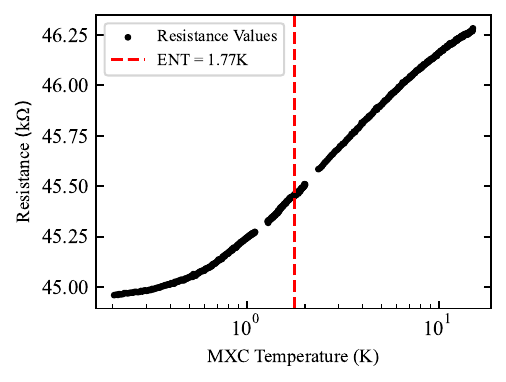}
    \caption{Resistance of $R_{M}$ as a function of temperature: The resistance of $R_{M}$ continues to change past the ENT of the base temperature noise measurement. This would suggest that $R_{M}$ is still thermalised to the MXC plate and hence would rule out that the saturation in the temperature of $R_{M}$ as the source of the extra noise.}
    \label{fig:R45k_R_vs_T}
\end{figure}

Preliminary noise measurements at a base temperature of $\mathrm{T = 15~mK}$ indicated that the measurement resistor exhibited an anomalously elevated noise floor, as depicted in Fig. \ref{fig:Initial_noise_measurement}.
Although this noise level is lower than that of the amplifier, it possesses an ENT of 1.77~K. To ascertain that this observation was not an artifact of the experimental configuration, noise measurements across the resistor were also conducted in an alternative section of the setup (R = 9~k$\mathrm{\Omega}$) and with a dead short of the measurement resistor, as illustrated in Fig. \ref{fig:Spectral_differences}.
Both of these values are below the noise level of the measurement resistor at base temperature. This indicates that this excess noise is related to the measurement resistor and not the set up. We consider two options for the existence of this unexpectedly high noise level.
\begin{enumerate}
    \item $R_{M}$ does not thermalise with the MXC and instead saturates at the ENT of 1.77K.
    \item The noise is caused by a combination of the base temperature thermal contribution and a constant offset from the properties of the resistor itself. 
\end{enumerate}
Figure \ref{fig:R45k_R_vs_T} illustrates the resistance of $R_{M}$ as a function of temperature.
This was determined through four-terminal lock-in measurements with an AC current excitation of 500~pA at a frequency of 66~Hz, with 11 values recorded per temperature step and subsequently averaged. The absence of a resistance plateau at the noise equivalent temperature of 1.77~K implies that at these temperatures, the resistor remains thermalised to the MXC rendering scenario 1 unlikely. The precise origin of the noise was challenging to determine due to the unknown construction of the resistor. However, literature has shown that non-thermal noise is prevalent in silicon-based resistors at low temperature\cite{deen1998low}. We consider this noise (denoted by $\mathrm{S_{Offset}(f)}$) to be temperature independent, thus treating it as an offset that can be subtracted from our correlated spectral values. To calculate this value, we take the a spectral measurement of the resistor at a well defined temperature, measure the resistance and then subtract the theoretically predicted Johnson-Nyquist value at this temperature. In practice this is performed at base temperature so to minimise the impact of error from the thermal noise contribution on the eventual noise offset value. The temperature at base was 15~mK based on the MXC thermometer reading. The theoretical base temperature contribution, $S_{T=15mK}$, was calculated using this temperature and a resistance reading. The total spectral density at base after sufficient cross correlation averages, $\langle S^{2}_{Base}(f) \rangle$, is given by, 
\begin{equation}
    \langle S^{2}_{Base}(f) \rangle = S^{2}_{Offset}(f) + S^{2}_{T=15mK}.
\end{equation}
For the average spectra at every temperature step, x, $\mathrm{\langle S_{MXC = x}(f) \rangle}$, we remove $\mathrm{S_{Offset}(f)}$, with the result $\mathrm{S_{T=x}(f)}$, 
\begin{equation}
    \langle S^{2}_{T=x}(f) \rangle = \langle S^{2}_{MXC = x}(f) \rangle - S^{2}_{Offset}(f), 
\end{equation}
which is then converted into a ENT by averaging the values in a range of white noise frequencies.
\section{Noise Temperature vs MXC temperature}
\begin{figure*}
    \centering
    \includegraphics{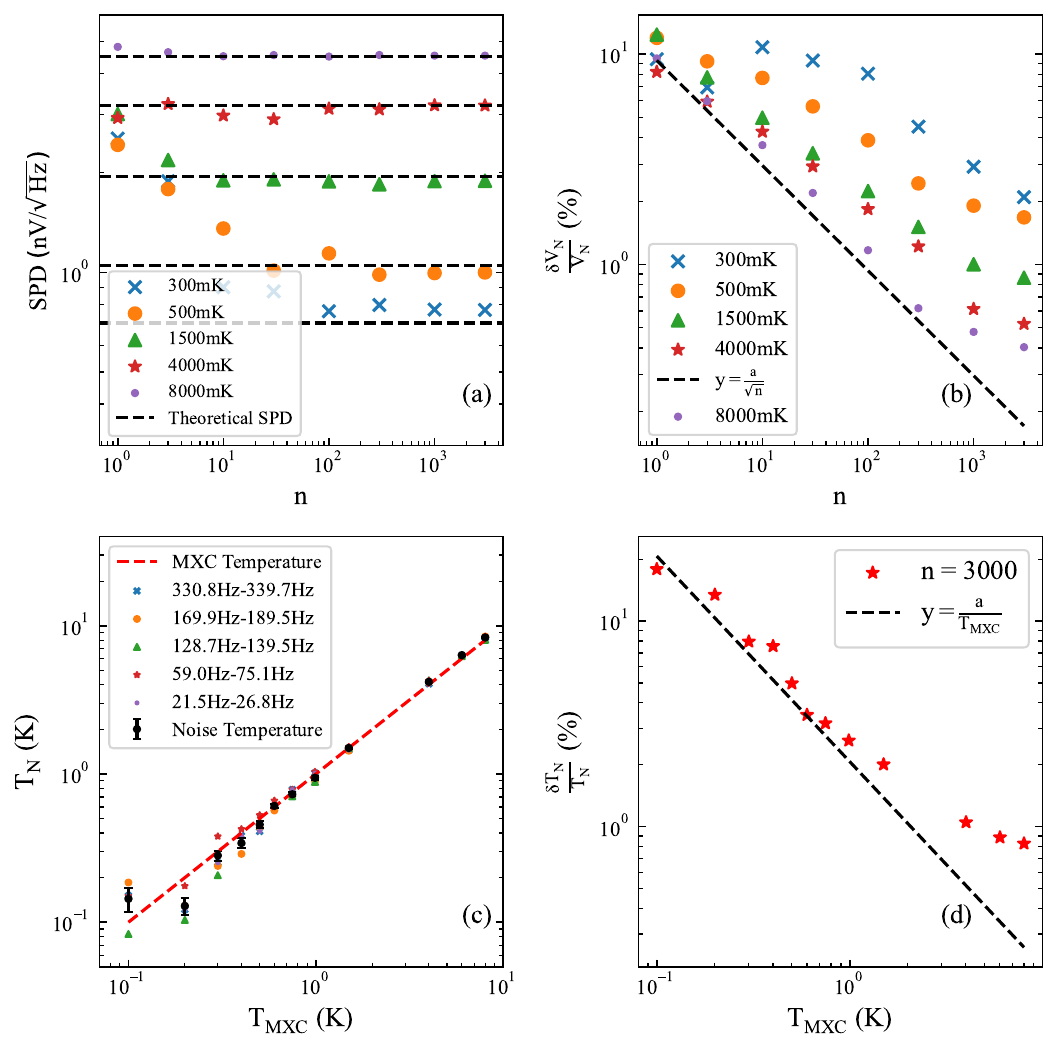}
    \caption{a) Measured Spectral density (SPD) as a function of the number of averages for various temperature ranges: At lower temperatures, increased averaging is required to sufficiently attenuate non-correlated noise and achieve the theoretical value. The theoretical SPD is derived from the Johnson-Nyquist formula, using the MXC temperature and a measured resistance value of $R_{M}$. b) Relative error in the SPD value as a function of the number of averages: The relative error is shown to be inversely proportional to the square root of the number of averages. This relationship begins to deteriorate at higher values of n, due to the inclusion of the error in the offset spectral density value, which remains constant. c) Noise temperature versus MXC temperature: Different-shaped markers represent the average values in various frequency ranges as indicated in the legend. The black points denote the average value across all frequency ranges at each temperature. This plot indicates that the noise temperature agrees in general with the MXC temperature. d) Relative error in temperature versus MXC temperature: This plot illustrates that the relative error in temperature is inversely proportional to the temperature, assuming that the Johnson-Nyquist SPD target noise is significantly lower than the SPD of the amplifier channel input. At elevated temperatures, this relationship deteriorates. }
    \label{fig:MXC_results_quad_plot}
\end{figure*}

In order to verify noise thermometry we compared the noise temperature of the resistor against the MXC thermometers. We increased MXC from 100~mK to 8~K in twelve steps. At each temperature step the resistance was measured to ensure its value was well known. At each step spectra were recorded, processed and averaged according to Sec. II. The results of this experiment are shown in Fig. \ref{fig:MXC_results_quad_plot}.
\begin{figure}
    \centering
    \includegraphics{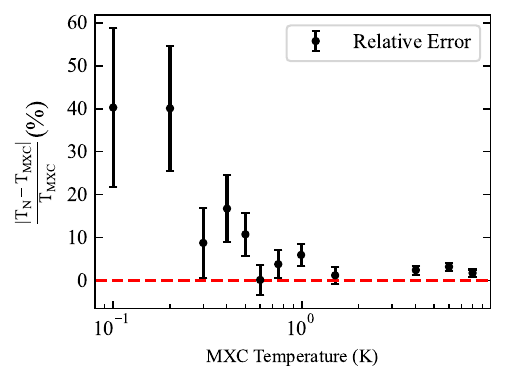}
    \caption{Relative error of the noise thermometry against the mixing chamber. The error bars are calculated from the standard error in the spectral measurements at these temperature values.}
    \label{fig:uncerts}
\end{figure}Further details on the agreement with the MXC temperature is provided in Fig. \ref{fig:uncerts}. Although the noise temperature generally aligns with the MXC temperature, it is only beyond 200~mK that the measured values begin to fall within the uncertainties of the MXC temperature. The uncertainty in the noise temperature also encompasses the uncertainty arising from the offset noise value. Therefore, employing a resistor without an offset will reduce the uncertainty across all temperature ranges. Additionally, due to the presence of noise offset, this method no longer qualifies as a form of primary thermometry. This is predicated on the assumption that the $R_{M}$ thermalises at the base temperature of 15~mK, an assumption that cannot be verified with certainty. Nonetheless, at temperatures exceeding 2~K, the impact of this assumption constitutes less than $\mathrm{1~\%}$ of the total value, rendering its effect almost negligible. \\ The principal limitation of this methodology is that the relative uncertainty is inversely proportional to the square root of the number of averages and inversely proportional to the temperature of the resistor. For instance, with 3000 averages, to achieve a relative uncertainty of less than 2$\mathrm{\%}$, the maximum measurable temperature is approximately 1.5 K. Alternatively, limiting measurement time to 1~minute requires 100 averages. For a relative uncertainty of 2$\mathrm{\%}$ this is limited to approximately  8~K or greater. Consequently, the primary constraint of this technique is the extended duration required for averaging to achieve lower temperatures and reduced relative uncertainties. The main solutions to this problem can be summarised as:
\begin{enumerate}
    \item Lower the input noise of the amplifiers.
    \item Decrease the time it takes to acquire one spectrum.
\end{enumerate}
One such alternative set up would be the use of cryo-amps. The lower noise floor of the amplifiers would allow a lower resistance value to be used; this would, in turn, increase the cutoff frequency of the RC circuit created by the resistor and parasitic wire capacitance. A higher cut-off frequency would allow for a higher sampling rate and reduce the time per spectral acquisition. A good example of accurate high frequency noise thermometry would be Crossno \textit{et al} \cite{crossno2015development}.
\section{Noise temperature vs power dissipation}
\begin{figure}
    \centering
    \includegraphics[width = 8.5cm]{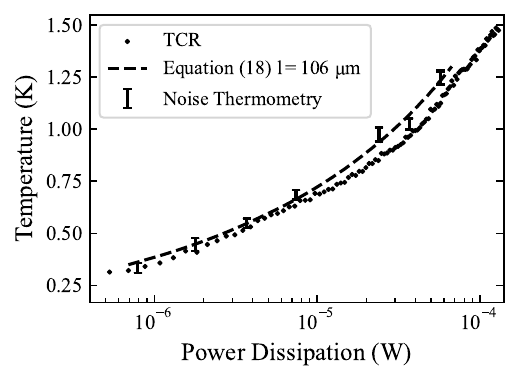}
    \caption{Temperature Coefficient of Resistance Thermometry (TCR) of $R_{M}$ and Noise Thermometry as a Function of Power: Noise thermometry agrees broadly with the TCR, but typically exceeds it. This could be explained through two scenarios. The first is that the baseline offset for the noise thermometry was slightly too large, possibly through the device not being properly thermalised to the MXC during its measurement. Secondly, if the TCR calibration was performed too quickly, it could also be suffering from a slight temperature offset. The dashed line fit fits Eq.\ref{eq:solution_heat_eq} to the noise data and gives an estimate of $l = 106~\mathrm{\mu m}$ (under the assumption that $l$ is either constant or weakly temperature dependent in the range of interest) which means that $Kn \approx 0.1$. This fit suggests a diffusive model is still appropriate. }
    \label{fig:T_Power}
\end{figure}

To substantiate noise thermometry, it was important to conduct \textit{in situ} testing by applying power within the IC while simultaneously measuring the temperature. This validation step is crucial, as the proximity of a power-dissipating element to $R_{M}$ could potentially induce cross-talk, thereby compromising the accuracy of noise measurements. To validate the noise thermometry, we utilised an alternative in-IC methodology, specifically the TCR technique that employs $R_{M}$. We used the temperature dependence characterised in Fig. \ref{fig:R45k_R_vs_T} as a calibration reference against the MXC. Initially, we tested TCR thermometry in relation to power dissipation ranging from 600~nW to $\mathrm{100~\mu W}$. This was to ensure the temperature was larger than 300~mK, at which temperatures the both the noise thermometer and TCR was considered to be accurate. At each increment, resistance measurements were taken and subsequently converted to temperature using the established calibration. Subsequently, to assess the noise thermometry, power was applied at seven levels, and at each level the spectra were processed and averaged in accordance with the methodology outlined in Sec. II. A baseline spectrum was recorded with the power resistor connected but without power dissipation to re-calibrate the offset spectral density. This was done to account for any potential cross-talk from the power resistor that might be present in all measurements. \\ The temperature inside the IC can be modelled in the idealised case that we consider the power source to be a point source at the centre of the top surface of the IC. We then consider the heat propagation to be radial from the heat source. We next have to consider, $l$, the mean free path of phonons within the silicon. This value is not well defined but can be large (on the order of centimeters \cite{swartz1989thermal}) in pure silicon at low temperatures. If $l$ approaches the characteristic length scale of the phonon container ($L$, in our case $2~\mathrm{mm}$) then the heat equation breaks down. This is because generating a temperature gradient requires multiple phonon scattering events throughout the length that is defined to have a gradient. In this case phonons are considered to behave ballistically, and the Boltzmann Transport Equation is more suited to describing the temperature profile \cite{wang2024modeling}. However, in ICs there are many places to provide scattering events. In the case of substrates with high dopant concentrations, this could be the dopant centers. Another place for scattering events is the top layer of the IC, which contains many different types of materials, which at low temperature can cause phonon mismatch at the boundaries and therefore scattering. The Knudsen number ($Kn$) helps to determine the boundary between ballistic and diffusive energy carrier propagation \cite{hahn2012heat} and is defined as \begin{equation}
    Kn = \frac{\Lambda}{L}.
\end{equation}
It gives the ratio between the characteristic distance an energy carrier travels ($\Lambda$) and the characteristic length scale of the system ($L$). In our case \begin{equation}
    Kn = \frac{l}{L}.
\end{equation}
If $Kn \le 0.1$, we can consider the system to act diffusively. If we use the diffusive model then the corresponding steady state heat equation is given by,
\begin{equation}
    \label{eq:Heat_eq}
    \frac{1}{r^{2}}\frac{\partial}{\partial r}\left( r^{2} \alpha T^{3} \frac{\partial T}{\partial r}\right) = 0,
\end{equation}
with the conductivity, $\kappa = \alpha T^{3}$. For silicon, this expression can be given by \cite{savin2006thermal,swartz1989thermal},
\begin{equation}
    \kappa(l) = \alpha T^{3} = 1200lT^{3}.
\end{equation}Equation \ref{eq:Heat_eq} does not contain a source term as its assumed all of the heat is dissipated back into the MXC. To solve it, several boundary conditions are given; the first is,
\begin{equation}
    r = R, \quad T = T_{Interposer}(Q)
\end{equation}
where $T_{Interposer}(Q)$ is the temperature of the silicon interposer at a given power. $R$ in our case is the distance from the power source to the silicon interposer, which is $1~\mathrm{mm}$. 
Interestingly, the temperature of the interposer greatly exceeds the copper holder and the MXC temperature (see Appendix D) due to the large boundary resistance at the copper - varnish - interposer boundary. Next,\begin{equation}
    r \rightarrow 0, \quad \alpha T^{3}\frac{\partial T}{\partial r} = \frac{Q}{2\pi r^{2}}.
\end{equation} This boundary condition states that the heat flux is equal to the total power dissipated ($Q$) divided by the area of hemisphere created by the radial movement of heat from the top surface. The solution to this equation is then,
\begin{equation}
    T = \left[\frac{2Q}{1200\pi l}\left(\frac{1}{r} - \frac{1}{R}\right)    + T^{4}_{Interposer}(Q)\right]^{\frac{1}{4}},
    \label{eq:solution_heat_eq}
\end{equation}
where $r=221 \mathrm{\mu m}$. The results of both TCR, noise thermometry and Eq. \ref{eq:solution_heat_eq} fitted for $l$ are illustrated in Fig. \ref{fig:T_Power}. Both data sets exhibit reasonable degree of agreement, thereby substantiating the efficacy of noise thermometry across a range of power dissipation levels. However, variation in the different frequency bins at each measurement point suggests that this technique would be improved by increasing the number of point averaged, to average out this variation. These findings imply that this methodology could be suitable for application within ICs containing multiple power-consuming devices.
\section{Conclusions}
We have established that noise thermometry exhibits broad agreement with other thermometry in the range 300~mK to 8~K. Furthermore, it shows substantial agreement the predicted temperature, when employed to determine the IC temperature corresponding to a specific power dissipation. This form of thermometry was made to be effective by the use of both cross-correlation and the subtraction of the offset noise calculated from base temperature measurements. This work suggests that with the addition of these techniques, voltage noise thermometry could be broadly applicable to ICs developed with industrial CMOS processes. However, the offset noise does precludes the noise thermometry in this paper from being a primary method of thermometry. Employing a resistor devoid of additional noise would reduce the measurement uncertainties and retain the technique as a primary thermometry method. Although this work demonstrates the increase in the uncertainty of measurements at lower temperatures and the large measurement time, we describe how the use of cold amplifiers could rectify both of these issues. Furthermore, as the direction of quantum computing moves towards temperatures closer to 1~K to access higher cooling power\cite{camenzind2022hole,yang2020operation,petit2018spin,petit2020universal}, this method will require less averaging to attenuate uncorrelated noise. This will mean faster measurements and lower uncertainties. In addition, the implementation of this technique is straightforward and requires only standard instruments that are typically available in academic or commercial laboratories. Given that the devices used are only resistors, this method significantly reduces the time and resources required for the design of ICs, facilitating the placement of thermometers at multiple locations within an IC. Finally, we suggest that this technique is an excellent candidate for performing in-IC temperature measurements to map the thermal profile of operational quantum-classical circuitry because of its dual primary and dissipationless nature.
\section{Acknowledgments}This research is supported by the European Union’s Horizon 2020 research and innovation programme (European Microkelvin Platform 824109). The interposer device fabrication has been done at HNF - Helmholz Nano Facility, Research Center Juelich GmbH \cite{albrecht2017hnf}. 




\section{Appendix}
\subsection{Uncertainty in Temperature}
The expectation relative error in the spectral measurement $\frac{\delta S}{S}$, the ratio of the expected standard deviation to the expected spectral value, is calculated from \cite{rubiola20061}. This is a function of the amplifier channels' power spectral density, $\mathrm{A^{2}}$, the target noise level, $\mathrm{X^{2}}$, and the number of averages, n, we can look at how the relative error changes with temperature within our measurements. As,
\begin{equation}
    \frac{\delta S}{S} = \frac{1}{X^{2}}\sqrt{\frac{\left(A^{2}+X^{2}\right)^{2}+X^{4}}{2n}},
\end{equation}
and we know that $S\propto T$ then it follows that in the case that the relative error in the resistance is small, that,
\begin{equation}
    \frac{\delta T}{T} \propto \frac{1}{X^{2}}\sqrt{\frac{\left(A^{2}+X^{2}\right)^{2}+X^{4}}{2n}}.
\end{equation}
In the low temperature case, where $X^{2} \ll A^{2}$, this expression simplifies further,
\begin{equation}
    \frac{\delta T}{T} \propto \frac{1}{\sqrt{2n}}\left(\frac{A^{2}}{X^{2}} +1\right).
\end{equation}
We know that in our case $X^{2} = 4k_{B}TR$ in a 1Hz bandwidth and that $A^{2}$ is constant with temperature. As such, we can say,
\begin{equation}
    \frac{\delta T}{T} \propto \frac{1}{T\sqrt{n}}.
\end{equation}
\nocite{*}
\subsection{Uncertainty in Noise Temperature}
Given the Johnson-Nyquist relationship between temperature and spectral density given in Eq. \ref{eq:Temperature_cal} the relative uncertainty in temperature, $\frac{\delta T}{T}$, is given by,
\begin{equation}
    \frac{\delta T}{T} = \sqrt{ \left(\frac{\delta S^{2}_{D}}{S^{2}_{D}}\right)^{2} + \left(\frac{\delta R}{R}\right)^{2} }.
\end{equation}
In this experiment, $\delta R$ is given by the standard error of the resistance measurements taken at every temperature step, while $R$ is given by the average value. For the spectral density, $\delta S^{2}_{D}$ is taken from the standard error of the values of the frequency values chosen to be averaged. However, because we have to remove an offset at each step, we need to take this into account. As 
\begin{equation}
    S^{2}_{D} = S^{2}_{M} - S^{2}_{offset},
\end{equation}
where $S^{2}_{M}$ is the measured spectral density and $S^{2}_{offset}$ is the offset spectral density. We calculate $S^{2}_{offset}$ by,
\begin{equation}
    S^{2}_{offset} = S^{2}_{Base} - S^{2}_{T=15mK}
\end{equation}
where $S^{2}_{Base}$ is the base temperature measurement and $S^{2}_{T=15mK}$ is the Johnson-Nyquist value of the spectral density of the resistor at 15mK, assuming the resistor is well thermalised with the MXC. The only error in $S^{2}_{T=15mK}$ is the error of the resistance at base in this case,
\begin{equation}
    \delta S^{2}_{offset} = \sqrt{\left(\delta S^{2}_{Base}\right)^{2} + \left(S^{2}_{T=15~mK} \frac{\delta R_{T=15~mK}}{R_{T=15~mK}}\right)^{2}}.
\end{equation}
Now we sum the errors of $S^{2}_{M}$ and $S^{2}_{offset}$ in quadrature,
\begin{equation}
    \delta S^{2}_{D} = \sqrt{\left(\delta S^{2}_{M}\right)^{2} + \left(\delta S^{2}_{offset}\right)^{2}}.
\end{equation}
\subsection{Calibration of Niobium Nitride Thermometers}
\begin{figure}
    \centering
    \includegraphics[width = \linewidth]{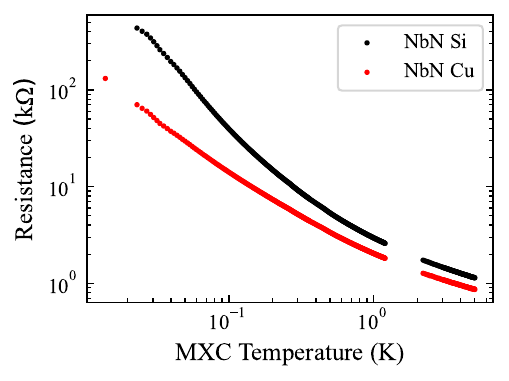}
    \caption{NbN calibration: The MXC temperature is given by a pre-calibrated ruthenium oxide thermometer placed on the fridge mixing chamber. }
    \label{fig:NbNs_cal}
\end{figure}

Niobium nitride thermometers were ideal candidates for monitoring both the copper and silicon interposer temperature due their excellent temperature sensitivity at milli-kevlin temperatures \cite{bourgeois2006liquid,nguyen2019niobium}. Both the interposer and copper thermometers had their resistance read out standard 4 point AC lock-in technique. The input current was 0.2~nA at 81~Hz. During the calibration, ten points were taken per temperature step, with a settling time of 3~s per point and a 3~s sweep delay. The first step to ensure adequate thermalisation was that the MXC temperature had to return two consecutive temperature readings within a defined threshold of the target temperature before it allowed the values to be taken. The temperature steps were also made to be small to further reduce the risk of poor thermalisation. From 16 mK to 360 mK the temperature step was 2 mK, from 360 mK onwards it was 5 mK. Results of the calibration are shown in Fig \ref{fig:NbNs_cal}. 
\subsection{Silicon Interposer, Copper and Mixing Chamber Temperature as a Function of Power}
\begin{figure}
    \centering
    \includegraphics[width=\linewidth]{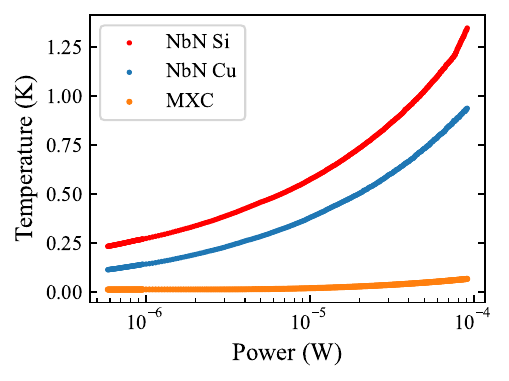}
    \caption{Temperature of the interposer thermometer (NbN Si) and the copper thermometer (NbN Si) and mixing chamber thermometer (MXC) as a function of power applied in the IC.}
   \label{fig:Si_Cu}
\end{figure}

Figure \ref{fig:Si_Cu}) demonstrates the temperature of several different thermometers at different locations on the setup. All thermometers were read out under the same conditions as their calibration. 

\bibliography{references}

\end{document}